\documentclass{article}

\usepackage{arxiv}

\usepackage[utf8]{inputenc} 
\usepackage[T1]{fontenc}    
\usepackage{hyperref}       
\usepackage{url}            
\usepackage{booktabs}       
\usepackage{amsfonts}       
\usepackage{nicefrac}       
\usepackage{microtype}      
\usepackage{lipsum}		
\usepackage{graphicx}
\usepackage{doi}
\usepackage{amsbsy}
\usepackage[square,sort,comma,numbers]{natbib}

\title{A new magneto-micropolar boundary layer model for liquid flows - effect of micromagnetorotation (MMR)}

\author{ \href{https://orcid.org/0000-0003-4614-2840}{\includegraphics[scale=0.06]{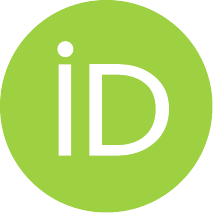}\hspace{1mm}Khan M.~Sabeel}\thanks{drmsabeel@gmail.com} \\
	Department of Mathematics\\
	Capital Univeristy of Science and Technology\\
	44000 Islamabad \\
	\texttt{muhammad.sabeel@cust.edu.pk} \\
	\And
	\href{https://orcid.org/0009-0002-9788-4713}{\includegraphics[scale=0.06]{orcid.pdf}\hspace{1mm}Isma~Hameed} \\
	Department of Mathematics\\
	Capital Univeristy of Science and Technology\\
	44000 Islamabad\\
	\texttt{ismakhan50@gmail.com} \\
}

\renewcommand{\shorttitle}{\textit{arXiv} Template}

\hypersetup{
pdftitle={A template for the arxiv style},
pdfsubject={q-bio.NC, q-bio.QM},
pdfauthor={David S.~Hippocampus, Elias D.~Striatum},
pdfkeywords={First keyword, Second keyword, More},
}

\begin{document}
\maketitle

\begin{abstract}
In this paper, we present a micropolar continuum model based on the theory of magneto-hydrodynamics. In particular, the effect of micromagnetorotation (MMR) is taken into account in the derivation of an initial-boundary value problem ($i$-$bvp$) within magneto-micorpolar flows. MMR is a phenomenon that is related to the micromotions of the magnetic liquid particles in the presence of externally applied magnetic field. In all previous investigations magnetization was supposed to be parallel to applied magnetic field therefore its effect in the lateral direction is neglected. This assumption is not correct in magnetic-micropolar flows. Since, magnetic-micropolar flows are an-isotropic in nature. Here, we present a model accounting for this MMR effect. The constitutive equation for the MMR is described and the governing system of flow dynamics is described in the form of PDEs. Boundary layer flow assumptions are used to derive an initial-boundary value problem in ODEs. As a consequence, two newly defined parameters arises that are related to the MMR. The effects of these parameters on the flow characteristics are investigated. The developed $i$-$bvp$ is solved through the shooting method using MATLAB routines. Effects of MMR are analyzed on the miro-rotational and hydrodynamic velocities profiles. Some interesting features of the flow are observed. Results are presented through graphs and discussed in detail. It is worth mentioning that the model presented is first of its kind in the literature and has a great potential in investigating boundary layer flows within micropolar continuum with other physical aspects of the flow pertinent to engineering and biomedical applications.
\end{abstract}


\section{Introduction}
\label{intro}
Boundary layer flows have been discussed in literature within the framework of micropolar continuum in different studies pertinent to various manufacturing applications. For instance,  in liquid films, cables, plastic films, cables, crystal making, the cooling process of metals, uninterrupted filament extrusion from tint, the winding roll, liquefied dynamics, and many other applications, see for instance \cite{Eringen1999, Eringen1966, Ariman1973, Jian-JunShu2008}. (crude oil extraction, food processing, extrusion of polymers, and syrup drugs.)\\

The micropolar fluid theory has wide application in the rheologic depiction of fluids for instance colloidal suspensions, liquid crystals, exotic lubricants and blood. Karvelas et al. \cite{KARVELAS2020} considered the blood as micropolar fluid and investigated the flow of blood inside a human carotid model. The work focused on the differences arising from the blood microstructure compared with a classical Newtonian fluid and found a significant decrease in the shear stress at the walls when the vortex viscosity and the microrotation increased. The combined effect of electrodynamics and magnetohydrodynamics of micropolar fluids was examined by Eringin \cite{Eringen1999}. MS Khan and Klaus Hackl \cite{MSKhan2021} presented the analytical and numerical aspects of with the help of relaxation energies.  microstructures' model within the framework of Cosserat continuum. They analysed that, in this framework the interaction energy potential enriches the material's free energy that accounts for the particles' counter rotations.   Recently, Saraswathy et al.\cite{Saraswathy2022} studied the non linear thermal radiation and viscous dissipation of MHD micropolar fluid through a porous channel and analyze that the spin gradient viscosity and vortex viscosity exhibits reverse phenomenon on microrotation profile. Almakki et al.\cite{Almakki2021} presented a micropolar nanofluid model for MHD unsteady flow in the presence of entropy generation through a streched surface on boundary layer.A non-Fick mass flux and non Fourier heat Cataeneo-Chritove model of micropolar MHD nanofluid flow through a radialy srteched disk was examined by Ayele Tulu \cite{Ayele2023}. A non linear model of PDEs for momentum, microrotation, thermal and concentration with boundary conditions and then solved numerically via spectral local linearization method. Reddy et al.\cite{Reddy2015} studied the stagnation point flow of MHD nanofluid flow over a streching sheet under the influence of magnetic induction. The influence of  MHD heat transfer on upper convected Maxwell micropolar flow with thermal radiation and joule heating was investegated by Khan et al.\cite{MSKhan2017}. They taken into account the microrotation of fluid particles and obtained angular-momentum balance equation and examined the microstructural parameters effects on microrotations and macroscopic  velocity profiles. Narayana et al. \cite{Narayana2013} discussed the radiation absorption and hall current effect on magnetohydrodynamic micropolar flow in a rotating reference frame and analyzed the magnetic field influence is normal to the porous surface which absorbs micropolar fluid with uniform suction velocity. B. Shankar Goud \cite{Goud2023} investegated the effect of heat generation and absorption in micropolar MHD flow past over a porous medium in the existence of porous suction and injection under the influence of thermal radiation and magnetic field. The obtained  non linear ODEs are  solved through RK-4 method along with shooting technique. A finite element study of MHD micropolar flow over a rectangural channel based upon itterative and numerical approached was investigated by MS khan et al. \cite{MSKhan2023}. They observed that the increase  in micropolar coupling parameter causes a decrease in microrotational velocity of particles as well as magnetic induction. MS khan and I. Hameed \cite{MSKhan2023} considered heat transfer analysis within the framework of MHD micropolar  flow using FreeFEM++.\\

In almost all the present studies in the literature within the context of micropolar boundary layer flows the effect of magnetization is not taken into account in direction lateral to applied magnetic field. However, the magnetization do effect \cite{SHIZAWA1986} the flow fields in the lateral direction. Here in this paper, we develop an $i$-$bvp$ based on the MMR phenonmenon within the theory of magneto-hydrodynamics. In this connection, the boundary layer flow approximations have been used. The developed $i$-$bvp$ has been investigated for the hydrodynamic flow characteristics for varying values of MMR. The rest of the paper is organized as follows: In section~\ref{MathForm}, the mathematical description of the flow dynamics is presented. In section~\ref{IBVPsec}, the derived initial-boundary value problem is described. In section~\ref{ResDis}, obtained results are presented and discussed. Finally, conclusions are drawn in section~\ref{conc}.

\section{Mathematical Model}\label{MathForm}
Consider an incompressible, laminar, two dimensional flow of MHD convective micropolar fluid under the application of an externally applied magnetic field. The governing flow equations along with the constitutive relation for magnetization are assumed as described in \cite{KhanIsma2023} and are stated as
 \begin{equation}\label{massconsv}
    \nabla\cdot\boldsymbol{U}=0,
\end{equation}
\begin{equation}
    \nabla \cdot \boldsymbol{H}=0,
\end{equation}
\begin{equation}
    \rho\left( \frac{d\boldsymbol{U}}{dt}+ \boldsymbol{U}\cdot \nabla \boldsymbol{U}\right) =-\nabla p + \eta \nabla^2 \boldsymbol{U} + 2\eta_1 \nabla \times (\boldsymbol{W}-\boldsymbol{w})+(\nabla \times \boldsymbol{H})\times \boldsymbol{B} +    
    (\boldsymbol{M} \cdot \nabla)\boldsymbol{H} + \boldsymbol{M} \times (\nabla \times \boldsymbol{H}),
    \end{equation}
\begin{equation}
    \rho \left(\frac{d\boldsymbol{H}}{dt}+ \boldsymbol{U} \cdot \nabla\boldsymbol{H} - \boldsymbol{H}\cdot\nabla \boldsymbol{U} \right) = \eta \nabla^2 \boldsymbol{H},
\end{equation}
\begin{equation}
    l \left(\frac{d\boldsymbol{W}}{dt}+ \boldsymbol{U} \cdot \nabla\boldsymbol{W}\right)=\gamma \nabla^2 \boldsymbol{W} + 4\eta_1 (\boldsymbol{w}-\boldsymbol{W}) + \boldsymbol{M}\times \boldsymbol{H},
\end{equation}
 \begin{equation}
    \boldsymbol{B}= \mu_0 \boldsymbol{H}+ \boldsymbol{M},
\end{equation}
\begin{equation}\label{magnet}
    \boldsymbol{M}= \frac{M_0(\boldsymbol{I}- \tau\boldsymbol{W}\cdot \epsilon)\cdot \boldsymbol{H}}{ \bar{H}}.
\end{equation}
where $\rho$ is the fluid density, $p$ is the pressure,  $j$ is the current density, $l$ is the moment of inertia, $\mu_0$ is the magnetic permeability, $\sigma$ is the electrical conductivity, $\boldsymbol{B}$ is magnetic induction vector and $\eta$, $\eta_1$, $\gamma$ are the shear viscosity, vortex viscosity and angular viscosity coefficients, respectively. The term $ \boldsymbol{M} \times \boldsymbol{H}$ is the micromagnetorotation(MMR) term, which specifies the influence of magnetization on microrotation. In the absence of electric field $\boldsymbol{E}$, and considering the assumptions used in \cite{KhanIsma2023} and the boundary layer approximations, the component form of the above model is reduced to
  \begin{equation}
      u\frac{\partial u}{\partial x}+ v\frac{\partial u}{\partial y}= \left(\frac{\mu+k}{\rho}\right) \frac{\partial^2 u}{\partial y^2}-\frac{\sigma}{\rho} \mu_0 H_2^2 u + \frac{k}{\rho} \frac{\partial N}{\partial y} + \frac{1}{\rho} \frac{M_0}{\bar{H}} \left(H_1 + \tau NH_2\right)\frac{\partial H_1}{\partial x} + \left(H_2- \tau N H_1\right) \frac{\partial H_2}{\partial x},
  \end{equation}
  \begin{equation}
      u\frac{\partial H_1}{\partial x}+v \frac{\partial H_1}{\partial y}= H_1 \frac{\partial u}{\partial x}+ H_2 \frac{\partial u}{\partial y}+ \mu_0 \frac{\partial^2 H_1}{\partial y^2},
  \end{equation}
  \begin{equation}
        u\frac{\partial N}{\partial x}+v \frac{\partial N}{\partial y}= \frac{\gamma^*}{\rho j}\frac{\partial^2 N}{\partial y^2} - \frac{k}{\rho j} \left(2N+ \frac{\partial u}{\partial y}\right)- M_0 H_0 \tau \bar{H} N.
  \end{equation}
\section{Initial-Boundary Value Problem}\label{IBVPsec}
Consider the boundary layer flow as shown in the Figure \ref{Fig1} below, the following boundary layer approximations are used.
\begin{figure}[h]
\centering
\includegraphics[width=10cm]{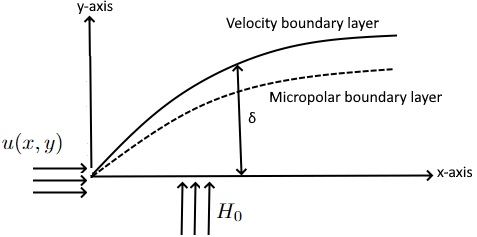}
\caption{Geometrical description of boundary layer along a surface stretched through fixed velocity.}
\label{Fig1}
\end{figure}
\begin{equation}\label{Transform1}
   u= cx f'(\eta), \quad v= - \sqrt{\nu c}f(\eta),
	\end{equation}
	where $\eta= \sqrt{\frac{c}{\nu}}y$,
\begin{equation}\label{Transform2}  
\quad N= cx\sqrt{\frac{c}{\nu}} g(\eta), \quad H_1= H_0 x h'(\eta),\end{equation}
and $H_2= -H_0 \sqrt{\frac{\nu}{c}} h(\eta)$.
\\
Subjected to the boundary conditions:
\begin{equation}
    u=cx, \quad v=0, \quad \frac{\partial{H_1}}{\partial y}= H_2 = 0, \quad N=-N_0 \frac{\partial u}{\partial y} \quad \textit{at} \quad y=0,
\end{equation}
\begin{equation}
    u=ax, \quad H_1= H_e(x)= H_0 x, \quad  N  \rightarrow 0, \quad \textit{as} \quad y \rightarrow \infty.
\end{equation}
The transformation in (\ref{Transform1})-(\ref{Transform2}) leads to the development of the following initial boundary value problem
\begin{equation}\label{IBVP1}
    (1 + K) f''' - \left( \frac{R_m}{R_e} \right) h^2 f' + K g' + \alpha h'^2 - \beta g h h' - f'^2 + f f'' = 0,
\end{equation}
\begin{equation}\label{IBVP2}
    \lambda h''' - hf'' + h'' f = 0,
\end{equation}
\begin{equation}\label{IBVP3}
    \left( 1 + \frac{K}{2} \right)g'' - \beta (2g+ f'')+ \displaystyle\left(\frac{\beta}{\alpha}\right) g - f' g + fg' = 0.
\end{equation}
The transformed boundary conditions at $\eta = 0$ are
 \begin{equation}\label{$i$-$bvp$4}
     f = 0, \quad f' = 1, \quad h = 0, \quad h'' = 0, \quad g = -N_0 f'',
 \end{equation}
and at $\eta \rightarrow \infty$ are
 \begin{equation}\label{$i$-$bvp$5}
     f'= \frac{a}{c}, \quad h'= 1, \quad g = 0.
 \end{equation}
  Here the parameters $K$, $R_m$, $R_e$, $d$, $\alpha$, $\lambda$, $\alpha$ and $\beta$ are dimensionless and are stated as follows:
$$
     K=\frac{k}{\mu}, \quad  \frac{R_m}{R_e}= \frac{\sigma H_0^2 \nu}{\rho c^2}, \quad R_m= \sigma \mu_0 v_0 L, \quad R_e= \frac{v_0 L}{\nu}, \quad  d = \frac{H_0^2}{c^2 l}, 
$$
$$
\quad \lambda = \frac{\mu_0}{\nu}, \quad  \left(1+\frac{k}{2}\right)=\frac{\gamma^*}{\rho j}, \quad \frac{\beta}{\alpha} =\frac{\tau M_0 \mu_0 \bar{H}}{c}, \quad \beta = \frac{\tau M_0 H_0^2}{\rho \bar{H}c}.
$$
The parameter of physical interest is the skin-friction coefficient defined as
\begin{equation}\label{SKF}
C_f = \frac{\tau_w}{\rho u_w^2},
\end{equation}
where $u_w = cx$ is the characteristic velocity and $\tau_w$ is the wall shear stress given by
\begin{equation}\label{wallSS}
\tau_w = \left[ (\mu + k) \frac{\partial u}{\partial y} + k N  \right]_{y = 0}.
\end{equation}
Substituting Eqns.~(\ref{Transform1}) and (\ref{Transform2}) into Eq.~(\ref{wallSS}), the skin-friction coefficient in Eq.~(\ref{SKF}) takes the following form
\begin{equation}\label{SFric}
C_f \sqrt{Re_x} = (1 + K) f''(0) + K g(0),
\end{equation}
where $Re_x = u_w x/\nu$ is the local Reynolds number.
\section{Results and Discussion} \label{ResDis}
\begin{figure}[h]
\center
\includegraphics[scale=0.65]{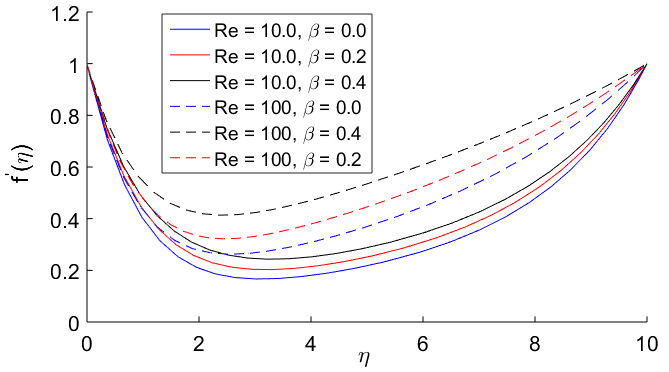}
\caption{Influence of MMR with varying Reynold number on microrotational velocity profile}\label{Fig2}
\end{figure}

\begin{figure}[h]
\center
\includegraphics[scale=0.65]{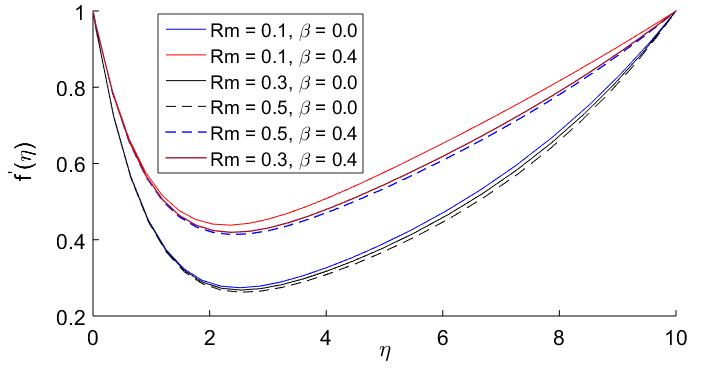}
\caption{Hydrodynamic velocity profile with varying magnetic Reynold number}\label{Fig3}
\end{figure}

\begin{figure}[h]
\center
\includegraphics[scale=0.65]{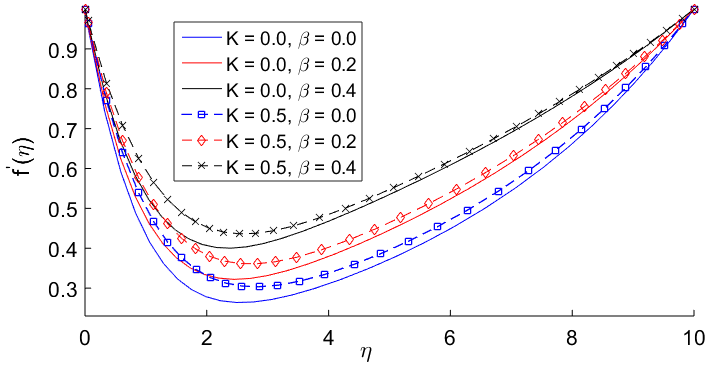}
\caption{Hydrodynamic velocity boundary layer profile with increasing MMR}\label{Fig4}
\end{figure}

\begin{figure}[h]
\center
\includegraphics[scale=0.65]{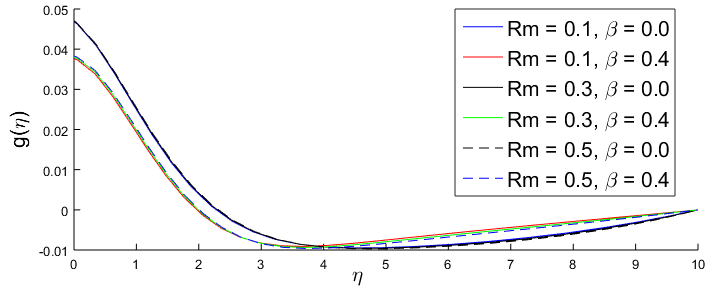}
\caption{Influence of $\lambda$ on macrorotational velocity profile in the absence and presence of MMR}\label{Fig5}
\end{figure}

\begin{figure}[h]
\center
\includegraphics[scale=0.65]{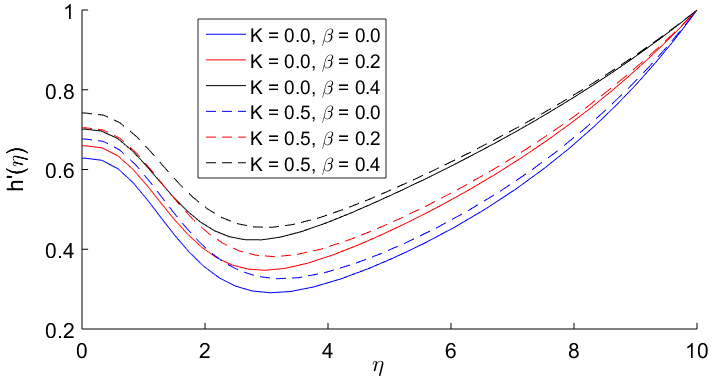}
\caption{Magnetic induciton $H_1$ profiles for varying values of MMR parameter.}\label{Fig6}
\end{figure}

In order to solve the initial boundary value problem in (\ref{IBVP1})-(\ref{$i$-$bvp$5}), we implement the shooting method. For this purpose, the $i$-$bvp$ is first converted to fist order ODE system. The shooting technique is taken together with Runge-Kutta $4^{th}$ order scheme. The graphical results have been shown for different values of material parameters to compute macrorotation and microrotation velocity profiles under the influence of MMR. In Figure~\ref{Fig2}, the hydrodynamic velocity boundary layer profile for different values of Reynolds number are shown with the varying values of magnetization parameter $\beta$. It can be seen that the hydrodynamic velocity profiles shows an increasing trend for higher Reynolds number with varying values of magnetization parameter. 
The influence of magnetic Reynolds number $R_m$ on hydrodynamic velocity boundary layer profile is depicted in Figure~\ref{Fig3} in the absence and presence of MMR parameter. It is noted that, the hydrodynamic velocity boundary layer profiles are lower in the absence of micromagnetorotation whereas, these profiles get higher in the presence of MMR. 
The comparison of hydrodynamic velocity boundary layer profiles is shown in Figure~\ref{Fig4} for different values of MMR. The behaviour of boundary layer profiles is examined in both the classical and micropolar flow description. The MMR is taken into account with varying magnitude of magnetization as shown in the figure. It is observed that the velocity boundary layer profiles are lower in case of classical fluid. 
In Figure~\ref{Fig5}, the influence of magnetic Reynolds number is seen on the microrotation velcoity profiles in the absence and presence of MMR. It is seen that in the presence of MMR the microrotation of boundary layer thickness is lower than in the one in the absence of MMR near the stretching plate. However, this behavior get reverse in the region away form the stretching plate. 
In Figure~\ref{Fig6}, the influence of micropolar constant on the development of magnetic induction profiels in the presence and absence of MMR. It is seen that the magnetic induction profiles increases with increasing value of the magnetization. Moreover, the magnetic induciton profiles in the case of micropolar fluid are higher than those in the case of classical fluid. 
In Table~\ref{SFVaryBetaStr}, the effect of micro-inertial coupling parameter $\beta^*$ is calculated on the skin-friction coefficient for both the classical and micropolar cases in the absence and presence of MMR term. To consider the effect of MMR term the values of the material parameter $\beta$ are chosen to be $0.3$ and $0.5$. It can be seen that in case of classical continuum the skin-friction coefficient does not changing significantly with increase in the value of micro-inertial coupling parameter. This is significantly due to non-involvement of the physical property of the micro-inertial coupling parameter in case of classical continuum. However, in case of micropolar continuum model with and without the inclusion of the MMR term, the skin-friction coefficient reduces with a decrease in value of micro-inertial coupling constant.
\begin{table}\centering
\caption{Skin-friction coefficient $C_f\sqrt{Re_x}$ with varying $\beta^*$ for different $\beta$ and $\lambda = 0.5.$}
\label{SFVaryBetaStr}       
\begin{tabular}{l lll l}
\noalign{\smallskip}\hline
&   & For $K = 0$\\
\hline\noalign{\smallskip}
$\beta^*$ & $\beta = 0$ & $\beta = 0.3$ & $\beta = 0.5$ \\
\noalign{\smallskip}\hline\noalign{\smallskip}
0.1 & -0.924142973799639 & -0.79100922788993 & -0.668524882690862 \\
0.5 & -0.924142977756218 & -0.789678769407542 & -0.664439355863969 \\
0.9 & -0.924142994623072 & -0.789573910431576 & -0.66347099715247 \\
1.3 & -0.924142996871294 & -0.789686923184627 & -0.663219124182537 \\
\noalign{\smallskip}\hline
&   & For $K = 1.0$\\
\hline\noalign{\smallskip}
$\beta^*$ & $\beta = 0$ & $\beta = 0.3$ & $\beta = 0.5$ \\
\noalign{\smallskip}\hline\noalign{\smallskip}
0.1 & -1.24396587733001 & -1.04161082758066 & -0.865391704012286 & \\
0.5 & -1.23512978324279 & -1.03358014099413 & -0.856071870942798 \\
0.9 & -1.22701806192418 & -1.02699203254663 & -0.849849209619443 \\
1.3 & -1.22036998621631 & -1.02182272852008 & -0.84534698464462\\
\noalign{\smallskip}\hline
\end{tabular}
\end{table}
\section{Conclusion}\label{conc}
So far, magnetization has been considered to be parallel to the externally applied magnetic field and thus ignored in all previous studies within micropolar MHD models. Here, we examine the impact of magnetization in lateral to applied magnetic field and study micromagnetorotation (MMR) on MHD micropolar flow. The aim is to analyze MMR effects on the boundary layer flow characteristics in the framework of micropolar continuum. MMR is the magnetization effect on micropolar flow associated with the implementation of applied magnetic field. The constitutive theory of MMR is used and governing flow dynamics is presented in the form of system of PDEs. The boundary layer approximations have been utilized to obtain an initial-boundary value problem which is afterwards solved numerically through shooting method. Impact of different associated dimensionless constants on microrotation and velocity fields has been investigated. The main points of this investigation are summarized as:\\
\begin{itemize}
    \item {The velcoity boundary layer thickness is reduced with increasing value of MMR.}\\
    \item {In the presence of MMR the microrotation boundary layer thickness is lower than in
the one in the absence of MMR near the stretching plate.}\\
    \item {The magnetic induction profiles increases with increasing value of the MMR}
    \item {The skin friction decreases with increasing magnitude of MMR.}
\end{itemize}
\newpage

\bibliographystyle{unsrt}
\bibliography{references}

\begin{thebibliography}{10}

\bibitem{Eringen1999}
A.~C. Eringen.
\newblock Microcontinuum field theories: I. foundations and solids.
\newblock {\em Springer Science and Business Media}, 1999.

\bibitem{Eringen1966}
A.~C. Eringen.
\newblock {\em J. Math. Mech.}, 16:1--18, 1966.

\bibitem{Ariman1973}
M.A.~Turk T.~Ariman and N.D. Sylvester.
\newblock {\em International Journal of Engineering Science}, 11:905--930,
  1973.

\bibitem{KARVELAS2020}
T.~Papathanasiou E.~Karvelas, G.~Sofiadis and I.~Sarris.
\newblock 5:125--137, 2020.

\bibitem{MSKhan2021}
M.~S. Khan and K.~Hackl.
\newblock {\em Trends in Applications of Mathematics to Mechanics}, pages
  103--125, 2021.

\bibitem{Saraswathy2022}
D.~Prakash M.~Saraswathy and P.~Durgaprasad.
\newblock {\em Mathematics}, 11:183, 2022.

\bibitem{Almakki2021}
H.~Mondal M.~Almakki and P.~Sibanda.
\newblock {\em International Journal of Ambient Energy}, 43:1--22, 2021.

\bibitem{Ayele2023}
T.~Ayele.
\newblock {\em Advances in Mathematical Physics}, (16), 2023.

\bibitem{Reddy2015}
N.~Sandeep C. S. K.~Raju J.~V. R.~Reddy, V.~Sugunamma and M.~J.Babu.
\newblock {\em Advanced in Science Engineering and Medicine}, 7:1--7, 2015.

\bibitem{MSKhan2017}
S.~Batool M.~S.~Khan, M.~Hammad and H.~Kaneez.
\newblock {\em The European Physical Journal Plus}, (132):158--182, 2017.

\bibitem{Narayana2013}
S.~Venkataramana P.V. S.~Narayana, B.~Venkateswarlu.
\newblock {\em Ain Shams Engineering Journal}, 4:843--854, 2013.

\bibitem{Goud2023}
B.~S. Goud.
\newblock {\em International Journal of Thermofluids}, pages 7--8, 2020.

\bibitem{MSKhan2023}
I.~Khan M.~S.~Khan, M. A.~Memon and S.~M. Eldin.
\newblock {\em Alexandria Engineering Journal}, 75:55--66, 2023.

\bibitem{SHIZAWA1986}
K.~Shizawa and T.~Tanahashi.
\newblock {\em Bull. JSME}, 29:2878–2884, 1986.

\bibitem{KhanIsma2023}
M.~S. Khan and I.~Hameed.
\newblock {\em Journal of Molecular Liquids}, 2023.

\end{thebibliography}

\end{document}